\documentclass[pre,showpacs,amsfonts,amssymb,eqsecnum,twocolumn,
floatfix,superscriptaddress,longbibliography]{revtex4-1}
\usepackage{amsmath}
\usepackage{graphicx}
\usepackage{bm}
\allowdisplaybreaks

\begin{document}

\title{Stochastic foundations in nonlinear density-regulation growth}
\author{Vicen\c{c} M\'{e}ndez}
\affiliation{Grup de F{\'i}sica Estad\'{i}stica.
Departament de F{\'i}sica. Facultat de Ci{\`e}ncies.
Edifici Cc. Universitat Aut\`{o}noma de Barcelona, 08193
Bellaterra (Barcelona) Spain}

\author{Michael Assaf}
\affiliation{Racah Institute of Physics, Hebrew University of Jerusalem,
Jerusalem 91904, Israel}

\author{Werner Horsthemke}
\affiliation{Department of Chemistry,
Southern Methodist University,
Dallas, Texas 75275-0314, USA}

\author{Daniel Campos}
\affiliation{Grup de F{\'i}sica Estad\'{i}stica.
Departament de F{\'i}sica. Facultat de Ci{\`e}ncies.
Edifici Cc. Universitat Aut\`{o}noma de Barcelona, 08193
Bellaterra (Barcelona) Spain}

\date{\today}

\begin{abstract}
In this work we construct individual-based models that give rise to the generalized logistic model at the mean-field deterministic level and that allow us to interpret the parameters of these models in terms of individual interactions. We also study the effect of internal fluctuations on the long-time dynamics for the different models that have been widely used in the literature, such as the theta-logistic and Savageau models. In particular, we determine the conditions for population extinction and calculate the mean time to extinction. If the population does not become extinct, we obtain analytical expressions for the population abundance distribution. Our theoretical results are based on WKB theory and the probability generating function formalism and are verified by numerical simulations.
\end{abstract}

\maketitle

\section{INTRODUCTION}
The main goal of population dynamics is to determine how the size of the population $n$ changes with time. It has been observed that the size $n(t)$ depends on the way the population grows when the size is small and declines when the size is large. In the small size limit, the simplest model predicts that the size  increases exponentially with time. This is equivalent to the assumption that the per capita growth rate is constant, $n^{-1}dn/dt=r$, where $r$ is the intrinsic per capita growth rate. This model yields an exponentially unlimited population growth, $n(t)=n(0)e^{rt}$. Although initial exponential growth occurs in many populations, it cannot extend to the whole population growth period. As the population grows, competition for resources among its individuals
and the presence of predators, parasites or competitors cause a reduction in the population growth rate. The relation between the per capita growth rate and the population size is one of the central issues in ecology. The classical model of population dynamics, the logistic equation, incorporates a density-dependent regulation that considers the simplest situation where the per capita growth rate declines linearly with population size, $n^{-1}dn/dt=r[1-(n/K)]$. Here $K$ is the carrying capacity of the population, i.e., it is the maximum sustainable population. The concept of carrying capacity in human populations has evolved to accommodate many resource limitations, such as available water, energy and other ecosystem goods and services \cite{Co95,Su13}.

The theta-logistic equation has been proposed as
a generalization of the logistic growth
 \cite{Gi73}. The per capita growth rate is
$n^{-1}dn/dt=r[1-(n/K)^\theta]$, where it is argued in
Ref.~\cite{Si05} that the exponent $\theta$ depends on the ways that animals interact at different densities. This model has been fitted to 1780 populations of birds, mammals, bony fishes, and insects \cite{Si05}, and both positive and negative values for $\theta$ were found. The negative values, however, have generated some controversy \cite{Re05,Ro09}. In fact, this parameter is very sensitive to measurement errors and environmental fluctuations \cite{Ba08}. It has been given a phenomenological interpretation to understand what ecological factors influence its value \cite{Th05}.
The basic Savageau model, where the per capita growth rate reads
\begin{equation}\label{eq:bSm}
n^{-1}dn/dt=rn^{\theta_1}[1-(n/K)^{\theta_2}],
\end{equation}
represents a more general model. It provides a more realistic framework in which to study density-dependent regulation, by allowing for simultaneous nonlinear effects of population size in birth and death rates \cite{Ro09,Ri15,Sa79,Ts02}.
A particular case of this model is the von Bertalanffy's equation \cite{Be57,Be60}, which has been applied as an ontogenetic growth model \cite{We01,Ho11} and in the context of the growth of cities \cite{Be07}.

All these models are deterministic and ignore fluctuations.
Some studies have considered the effect of external fluctuations on the theta-logistic equation by assuming ad-hoc Gaussian white noise \cite{LiWa12}. Yet, internal or demographic fluctuations,
caused by the discreteness of individuals and the stochastic nature of their interactions, are known to be very important in the case of small population sizes \cite{NiGu03,GoRi74}.

Some authors have investigated demographic noise in those models, by considering single-step birth-death processes that give rise to linear \cite{Na01,Ov01} and nonlinear density-regulation growth in the mean-field limit \cite{MaKiPa98,BhBaRa16}. However, they introduce
the density-dependent birth-and-death rates in an \textit{ad hoc} manner.
Furthermore, while the quasistationary distribution (QSD) has been analytically \cite{MaKiPa98} or numerically \cite{Ga90} calculated in some particular cases, the mean time to extinction has not been obtained for these models.

We investigate the dynamics of a single species population influenced by demographic fluctuations in the most general case. To this end, we adopt a general multi-step reaction scheme. This framework allows us to define the ecological system in terms of the events that govern the dynamics of the system at the individual level. In particular, instead of postulating it phenomenologically, as in Refs.~\cite{MaKiPa98,BhBaRa16}, we derive the generalized growth equation analytically as the mean-field limit of these individual-based models. Our goal is twofold: (i) To provide a clear physical meaning for the parameters of the mean-field deterministic growth equations in terms of the rates of the processes occurring in an individual-based description. (ii) To go beyond the mean-field description and to elucidate the effects of demographic noise on the population dynamics. Specifically, we determine the conditions for extinction or persistence. For the former, we employ the WKB theory~\cite{Dykman1994lfo} to obtain analytical expressions for
the QSD and the mean time to extinction (MTE). For the latter, we employ the probability generating function formalism to obtain analytical expressions for
the stationary population abundance.
Our theoretical results are verified with numerical simulations.

The paper is organized as follows. We introduce the individual-based description in Sec.~\ref{sec:individual}, derive the corresponding mean-field equation, and establish the connection between the rates of the microscopic processes and the parameters of the deterministic growth equation, achieving our first goal. In the following sections, we focus on our second goal. In Sec.~\ref{sec:twoa}, we determine the conditions for extinction due to demographic noise and
obtain analytical expressions for the QSD and MTE for
the generalized logistic model. For the case of population persistence, we derive the analytical expression for the population abundance distribution in Sec.~\ref{sec:abund}. We extend our individual-based description in Sec.~\ref{sec:three} to include a natural death process and investigate its effect on the MTE. In Sec.~\ref{sec:doom} we provide results for doomsday scenarios, i.e., runaway systems. We discuss our results and their implications in Sec.~\ref{sec:concl}.

\section{Stochastic description of nonlinear density-regulation models}\label{sec:individual}
We start from the general birth-and-death processes
\begin{subequations}\label{eq:1}
\begin{align}
b\text{A} & \xrightarrow{\lambda}  (b+a)\text{A},\\
c\text{A} & \xrightarrow{\mu}  (c-d)\text{A},
\end{align}
\end{subequations}
where $a$, $b$, $c$ and $d$ are positive integers and $c\geq d$. These interactions imply that $b$ individuals are necessary to produce $a$ new individuals at rate $\lambda$, and $c$ individuals fight or compete
to remove $d$ individuals at rate $\mu$.
We make the standard assumption that the reaction scheme \eqref{eq:1}
defines a Markovian birth-and-death process and employ
the Master equation, also known as the forward Kolmogorov equation,
to describe the temporal evolution of $P(n,t)$, the probability of
having $n$ individuals at time $t$ \cite{Ga90}:
\begin{equation}
\frac{\partial P(n,t)}{\partial t}=\sum_{r}\left[W(n-r,r)P(n-r,t)-W(n,r)P(n,t)\right].\label{eq:me}
\end{equation}
Here, $W(n,r)$ are the transition rates between the states with $n$
and $n+r$ individuals, and $r=\{r_{1},r_{2}\}=\{a,-d\}$ are the
transition increments. The transition rates corresponding to each
reaction, $W(n,r)$, are obtained from the reaction kinetics and for
\eqref{eq:1} read:
\begin{equation}
W(n,a)=\frac{\lambda}{b!}\frac{n!}{(n-b)!},\; W(n,-d)=\frac{\mu}{c!}\frac{n!}{(n-c)!}.\label{eq:tr}
\end{equation}
Substituting (\ref{eq:tr}) into (\ref{eq:me}), we find
\begin{multline}
\frac{\partial P(n,t)}{\partial t}=\frac{\lambda}{b!}\frac{(n-a)!}{(n-a-b)!}P(n-a,t)\\
+\frac{\mu}{c!}\frac{(n+d)!}{(n+d-c)!}P(n+d,t)\\
-\left[\frac{\lambda}{b!}\frac{n!}{(n-b)!}+\frac{\mu}{c!}\frac{n!}{(n-c)!}\right]P(n,t),\label{eq:me2}
\end{multline}
where it is understood that $P(n<0,t)=0$.
For $P(0,t)$, the master equation is
\begin{equation}
\frac{\partial P(0,t)}{\partial t}=\sum_{r<0} W(n=-r,r)P(-r,t).
\label{eq:P0}
\end{equation}

A population that undergoes the birth-and-death processes described by (\ref{eq:1}) can either become extinct, when $c=d$, or reach a nontrivial stationary distribution, when $d<c$. In the former case, we assume that prior to extinction the system first reaches a long-lived metastable state, whose shape is given by the QSD centered about the nontrivial steady state of the deterministic mean-field equation, see below.
(The concept of the QSD was introduced in Ref.\
\cite{DaSe65}.)
After an exponentially long waiting time, the population reaches the absorbing state at $n=0$ and becomes extinct. Importantly, once the system has settled into the long-lived metastable state, the dynamics is governed by a single time exponent. The time-dependent probability distribution function (PDF) satisfies $P(n>0,t)=P(n)\, e^{-t/\tau}$, while the extinction probability satisfies $P(0,t)=1-e^{-t/\tau}$, where $\tau$ is the MTE \cite{Dykman1994lfo,Assaf2006stm,assaf2007spectral,kessler2007extinction,meerson2008noise,escudero2009switching,Assaf2010ems,AssafMeerson2017}.
As a result, the MTE $\tau$ generally satisfies~\cite{Assaf2010ems}
\begin{equation}
\tau^ {-1}=\sum_{r<0} W(n=-r,r)P(-r),
\label{eq:met1}
\end{equation}
where $P(-r)$ has to be determined by matching the QSD, valid for large $n$, with a recursive solution valid for small values of $n$
\cite{Assaf2010ems}. The inverse of $\tau$ represents the population's extinction risk, the rate at which extinction may occur within a given period of time.

To address our first goal, we derive the mean-field dynamics from the master equation.
Multiplying \eqref{eq:me2} by $n$ and summing over $n$, we obtain
the equation for the mean number of individuals. Applying the mean-field
approximation $\left\langle n^{k}\right\rangle \simeq\left\langle n\right\rangle ^{k}$,
which holds if the typical population size is large,
we arrive at the mean-field equation for the population growth,
\begin{equation}
\frac{dn}{dt}=\frac{\lambda a}{b!}n^{b}-\frac{\mu d}{c!}n^{c},\label{eq:mfe1}
\end{equation}
known in the ecological literature as the \textit{basic Savageau model} \cite{Sa79}.
This achieves our first goal.
Comparing the equation for the Savageau model \eqref{eq:bSm} with the mean-field limit of the individual based description \eqref{eq:mfe1}
allows us to interpret the parameters $\theta_{1}$, $\theta_{2}$, and $K$ in terms of the basic parameters of the individual interactions, $a$, $b$, $c$, $d$, $\lambda$, and $\mu$. Explicit expressions are provided below.

\section{Stochastic dynamics of the theta-logistic equation. Two-reactions model}\label{sec:two}

We now consider a specific example of the reaction scheme given by Eq.~(\ref{eq:1}), namely the so-called theta-logistic equation,
\begin{subequations}\label{eq:rec1}
\begin{align}
\text{A} & \xrightarrow{\lambda}  2\text{A},\\
c\text{A} & \xrightarrow{\mu} (c-d)\text{A},
\end{align}
\end{subequations}
where $d\leq c$, and we have set $a=b=1$ in Eq.~(\ref{eq:1}).
Here, the birth process consists of an individual that produces a new offspring
with constant rate $\lambda$. The death process destroys $d$ individuals
as a consequence of competition among $c$ individuals. As stated above, extinction of the population can only occur when $d=c$, in which case the death process, given by (\ref{eq:rec1}), reads $c\text{A}\xrightarrow{\mu}\emptyset.$

The transition probabilities follow from (\ref{eq:tr}),
\begin{equation}
W(n,1)=\lambda n,\quad W(n,-d)=\frac{\mu}{c!}\frac{n!}{(n-c)!}.\label{eq:trc1}
\end{equation}
The master equation corresponding to the individual interactions (\ref{eq:rec1})
reads
\begin{multline}
\frac{\partial P(n,t)}{\partial t}=\lambda(n-1)P(n-1,t)\\
+\frac{\mu}{c!}\frac{(n+d)!}{(n+d-c)!}P(n+d,t)\\
-\left[\lambda n+\frac{\mu}{c!}\frac{n!}{(n-c)!}\right]P(n,t).\label{eq:mec1}
\end{multline}

As stated above, the mean-field dynamics can be found by multiplying the master equation~(\ref{eq:mec1}) by $n$ and summing over $n$. This
yields the theta-logistic equation \cite{Gi73} as a mean-field description of
the system (\ref{eq:rec1}),
\begin{equation}
\frac{dn}{dt}=rn\left(1-\frac{n^{c-1}}{K^{c-1}}\right).\label{eq:tl}
\end{equation}
The intrinsic growth rate and carrying capacity are defined
in terms of the parameters related to the individual interactions,
\begin{equation}
r=\lambda,\quad K=\left(\frac{c!\lambda}{\mu d}\right)^{\frac{1}{c-1}}.
\label{rK}
\end{equation}

This result contributes to our first goal. We have provided a clear meaning
of
the exponent $\theta=c-1$ of the nonlinear regulation term in the
theta-logistic equation. It corresponds to
the number of individuals involved in
the competition or death process minus one. In addition, we have obtained an expression for the carrying capacity $K$ in terms of the parameters of the individual-based model \eqref{eq:rec1}.

The deterministic behavior of the population is given by the mean-field theta-logistic
equation (\ref{eq:tl}). The state $n=0$ is an unstable state, while the
state $n_{*}=K$ is a stable
state. Further, (\ref{eq:tl}) can be solved exactly,
\begin{equation}
n(t)=\frac{K}{\left\{ 1-\left[1-\left(\frac{K}{n(0)}\right)^{c-1}\right]e^{-r(c-1)t}\right\} ^{1/(c-1)}}.
\end{equation}

The mean-field dynamics ignores demographic fluctuations originating in
the discreteness of individuals and the stochasticity of
the birth-and-death processes. Fluctuations cause extinction
of the population when $d=c$
and give rise to a
stationary probability distribution function for $d<c$.

To address our second goal, we determine the QSD of the long-lived metastable state and the MTE for the case of $d=c$ (extinction) and the population
abundance distribution for $d<c$ (persistence).
Note that these two, qualitatively very different, regimes of the underlying microscopic dynamics
are both described by the same mean-field equation,
namely Eq.\  (\ref{eq:tl}).

\subsection{Quasi-stationary distribution and mean time to extinction}\label{sec:twoa}
In this subsection we consider the case of $c=d$ and calculate the QSD and MTE employing the WKB theory~\cite{Dykman1994lfo}.
It is convenient to rescale time $t\to \lambda t$ and introduce the rescaled population
number density $q=n/N$, where
\begin{equation}
N=\left(\frac{\lambda}{\mu}\right)^{\frac{1}{c-1}}
\label{eq:N1}
\end{equation}
denotes the typical population size in the long-lived (meta)stable state. Henceforth we will assume that $N\gg 1$.

To determine the QSD and MTE, we follow the general formalism outlined in the papers by Escudero and Kamenev~\cite{escudero2009switching} and  Assaf and Meerson~\cite{Assaf2010ems}, and write the transition rates, for $N\gg 1$, as
\begin{equation}
W(Nq,r)=Nw_{r}(q)+u_{r}(q)+O(N^{-1}).\label{eq:exp}
\end{equation}
In the case of the theta-logistic model we have
\begin{align}
w_{a}&=q, \quad u_{a}=0,\label{eq:tr1c1}
\\
w_{-d}&=\frac{q^{c}}{c!}, \quad u_{-d}=-\frac{q^{c-1}}{2(c-2)!}.\label{eq:tr2c1}
\end{align}
Here $w_{r}$ and $u_{r}$ are $O(1)$. A
 necessary condition for extinction is that  $w_{r}(0)=u_{r}(0)=0$,
for any $r$, which indicates that the extinction state is an absorbing state. Note that the system can reach this state only when $d=c$ as
shown below.

\subsubsection{Quasi-stationary distribution}
The WKB approach employs the assumption
that for $N\gg 1$ the probability for rare events, such as extinction, to occur
lies in the tail of the distribution and falls away steeply from the
steady state. Substituting the WKB ansatz for the QSD, $P(n)\equiv P(Nq)$\cite{Dykman1994lfo,Assaf2006stm,assaf2007spectral,kessler2007extinction,meerson2008noise,escudero2009switching,Assaf2010ems,AssafMeerson2017},
\begin{equation}
P_{\text{WKB}}(q)\equiv Ae^{-NS(q)-S_{1}(q)},
\label{Pwkbmain}
\end{equation}
into the master equation (\ref{eq:mec1}), the functions $S(q)$ and $S_1(q)$ can be calculated order by order for $N\gg 1$, see the Appendix for the detailed calculations. Doing so, we obtain
\begin{equation}
P_{\text{WKB}}(q)=\sqrt{\frac{S''(q_{*})}{2\pi N}}\frac{q_{*}}{q}e^{N[S(q_{*})-S(q)]}e^{\phi(q_{*})-\phi(q)},\label{eq:QSD}
\end{equation}
where the action, $S(q)$, is given by Eq.~(\ref{eq:sq}), $\phi(q)$ is given by Eqs.~(\ref{eq:f1}) and (\ref{Lambda}), and $q_{*}=n_{*}/N=[(c-1)!]^{1/(c-1)}$.

The WKB solution~(\ref{eq:QSD}) is valid as long as $n\gg 1$ or $q\gg N^{-1}$. In order to find the MTE, see below, we have to match the WKB solution (\ref{eq:QSD}) to a recursive solution for the QSD valid for small $n$~\cite{Assaf2010ems}.

The recursive solution, which is the left tail of the QSD for $n\ll N$, i.e., sufficiently far from the mean-field stable fixed point, can be found by setting $\partial_{t}P(n,t)\simeq 0$ in the master equation (\ref{eq:mec1}).  To solve the resulting difference
equation recursively, we note that the QSD is rapidly growing for
small $n$. As a result, and since $\lambda\gg\mu$, we can neglect
the $\lambda$ term proportional to $P(n-1)$ compared to $P(n)$
and the $\mu$ term proportional to $P(n)$ compared to $P(n+d)$~\cite{kessler2007extinction}.
Denoting $\Omega=\lambda/\mu\gg1$, we obtain for $n\ll N$,
\begin{align}\label{eq:ps}
P(n+d)&=\frac{\Omega c!\, n(n+d-c)!}{(n+d)!}P(n)\\
&=\frac{\Omega c!\, nP(n)}{(n+d)(n+d-1)\cdots(n+d-c+1)}.\nonumber
\end{align}
Finally, as we are dealing with the case of extinction,
$d=c$, the solution of this recursion equation is given by
\begin{equation}
P(n)=\frac{[c\,c!\Omega]^{n/c}\,\Gamma(1+n/c)}{\Omega n^{2}\Gamma(n)}P(c).\label{recursivePDF}
\end{equation}
It can be shown that the applicability of this solution requires $n\ll N$~\cite{kessler2007extinction}.
Equations (\ref{eq:QSD}) and (\ref{recursivePDF}) constitute the complete QSD for $n>0$.

\subsubsection{Mean time to extinction}
To find the MTE, we set $c=d$ and use Eqs.~(\ref{eq:met1}) and (\ref{eq:trc1}). This yields $\tau^{-1}=W(n=d,-d)P(c)=\mu P(c)$.
To find $P(c)$, one needs to match the WKB, Eq.~(\ref{eq:QSD}), and recursive, Eq.~(\ref{recursivePDF}), solutions of the QSD
\cite{Assaf2010ems}. To match these solutions in their joint region of validity, $1\ll n\ll N$, we need to find the $n\ll N$ or $q\ll 1$ asymptote of the WKB solution (\ref{eq:QSD}).

We begin by calculating the $q\ll 1$ asymptote of $S(q)$. In the case of the theta-logistic model, Eqs.~(\ref{eq:rec1}), the Hamiltonian Eq.~(\ref{eq:H}) becomes
\begin{equation}
H(q,p)=q\left(e^{p}-1\right)+\frac{q^{c}}{c!}\left(e^{-dp}-1\right).\label{eq:Htheta}
\end{equation}
To find the action, one needs to find the non-trivial zero-energy trajectory of $H$, the optimal path to extinction, see Appendix. Here, the optimal path reads
\begin{equation}
q_{a}(p)=\left[c!\frac{e^{dp}\left(e^{p}-1\right)}{e^{dp}-1}\right]^{\frac{1}{c-1}}.\label{eq:qa}
\end{equation}
The action can be calculated by integrating over this trajectory, $S(q)=\int p_a(q)dq$, where $p_a(q)$ can be found from (\ref{eq:qa}) by solving $q_a(p)=q$. In the limit $q\to 0$, we have $p\to-\infty$.
Defining  $\epsilon=e^{p}\ll 1$,  we have $q\simeq(c!)^{1/(c-1)}\epsilon^{d/(c-1)}$ in the leading order in $\epsilon$. Therefore, for $q\ll 1$, the action function satisfies
\begin{equation}
S(q)\simeq\frac{c-1}{d}q\left[\ln\left(\frac{q}{(c!)^{1/(c-1)}}\right)-1\right]+O(q^{2}).\label{eq:Sq}
\end{equation}
As a result, the $q\ll 1$ asymptote of the WKB solution reads
\begin{multline}
P_{\text{WKB}}(n)\simeq \sqrt{\frac{S''(q_{*})}{2\pi N}}\frac{Nq_{*}}{n}e^{-N\Delta S-\Delta\phi}\\
\times e^{(c-1)\frac{n}{c}-(c-1)\frac{n}{c}\ln\left[n/\left(N^{c-1}c!\right)^{1/(c-1)}\right]}.\label{eq:pwkb2}
\end{multline}
Here, using Eq.~(\ref{eq:sq}),
\begin{align}
\Delta S&=S(q=0)-S(q_{*})=\int_{q_{*}}^{0}p_{a}(q)dq=\int_{-\infty}^{0}q_{a}(p)dp\nonumber\\
&=(c!)^{\frac{1}{c-1}}\int_{0}^{1}\left(z\frac{1-z}{1-z^{c}}\right)^{\frac{1}{c-1}}dz,
\label{DeltaS}
\end{align}
is the entropic barrier to extinction, where we have introduced the new variable $z=e^p$, and from Eq.~(\ref{eq:f1}),
\begin{equation}
\Delta\phi  =  \int_{0}^{-\infty}\frac{dq_{a}(p)}{dp}
\Lambda (q=q_a(p),p )dp = \ln\left(\frac{\sqrt{2}c}{\sqrt{c+1}}\right).
\label{eq:df}
\end{equation}
Note that the result given in (\ref{eq:df}) only holds
for $d=c$. For $d<c$, a general calculation shows that $\Delta\phi$ diverges, indicating that the extinction rate vanishes.

On the other hand, to calculate
 the $n\gg 1$ or $q\gg N^{-1}$ asymptote of the recursive solution (\ref{recursivePDF}), we use the Stirling approximation $\Gamma(1+n)\simeq\sqrt{2\pi n}(n/e)^{n}$ in Eq.~(\ref{recursivePDF}), which now becomes
\begin{equation}
P(n)\simeq P(c)\frac{e^{\frac{n}{c}\left[\ln(c!\Omega)-(c-1)(\ln n-1)\right]}}{\sqrt{c}n\Omega},\label{recasym}
\end{equation}
valid for $1\ll n\ll N$.

Matching the asymptotes (\ref{eq:pwkb2}) and (\ref{recasym})
in their joint region of validity, $1\ll n \ll N$, we find
\begin{equation}
\tau=\sqrt{\frac{2\pi c}{c-1}}\frac{e^{N\Delta S}}{\mu N^{c-1/2}[(c-1)!]^{1/[2(c-1)]}},
\label{eq:met0}
\end{equation}
where $\Delta S$ is given by Eq.~(\ref{DeltaS}).

\begin{figure}[htbp]
\includegraphics[width=\hsize]{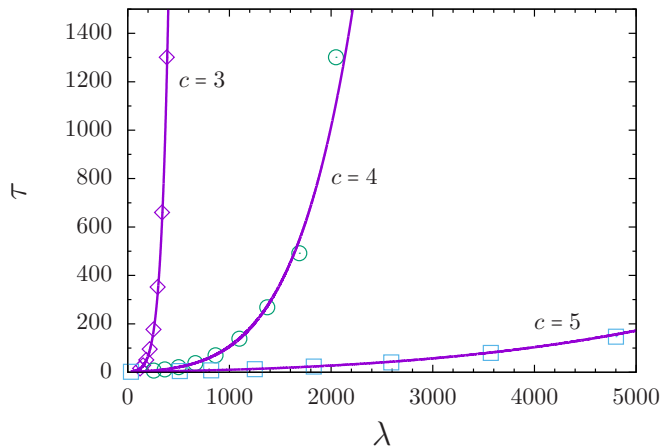}
\caption{MTE vs $\lambda$ for the processes given by $\text{A}  \xrightarrow{\lambda}  2\text{A}$ and $c\text{A}  \xrightarrow{\mu}  \varnothing$, for $\mu=3$. The value of $N$ is given by (\ref{eq:N1}). Symbols correspond to numerical simulations performed up to time $10^8$ and averaging over $10^4$ realizations. Solid curves for $c=3,4,5$ are respectively obtained from Eqs.~(\ref{met2}), (\ref{met3}), and Eq.~(\ref{eq:met0}) with $c=5$.}
\label{fig:f1}
\end{figure}

This analytical formula
is one of our main results. It displays explicitly how the parameters characterizing the individual interactions affect the MTE.
It generalizes previous results obtained for specific values of $c$
\cite{assaf2007spectral,assaf2010large}. For example,
for $c=2$, using Eq.~(\ref{eq:met0}) we find
\begin{equation}
\tau_{c=2}=\frac{2\sqrt{\pi}}{\mu\, N^{3/2}}e^{2N(1-\ln2)}.
\label{met1}
\end{equation}
For $c=3$ we find
\begin{equation}
\tau_{c=3}=\frac{\sqrt{3\pi}}{\mu\,2^{1/4}N^{5/2}}e^{1.183N},
\label{met2}
\end{equation}
and for $c=4$,
\begin{equation}
\tau_{c=4}=\frac{2^{4/3}\sqrt{\pi}}{\mu\, 3^{2/3}N^{7/2}}e^{1.699N}.
\label{met3}
\end{equation}
The results for $c=2,3$ are in agreement with previous calculations
\cite{assaf2007spectral,assaf2010large}. The number of individuals involved in the competition or death process, $c$, strongly affects
the value of the MTE, see Fig.~\ref{fig:f1}. When the birth rate $\lambda$ increases, the MTE also increases. This means that, as expected, the risk of extinction decreases when the birth rate increases. On the other hand, for a given birth rate, increasing the number of individuals participating in the competition or death process decreases the MTE and increases the extinction risk of the population, see Fig.~\ref{fig:f1}. Note that the dependence of the MTE on $c$ is very strong.

%%%%%%%%%%
\begin{figure}[htbp]
	\includegraphics[width=\hsize]{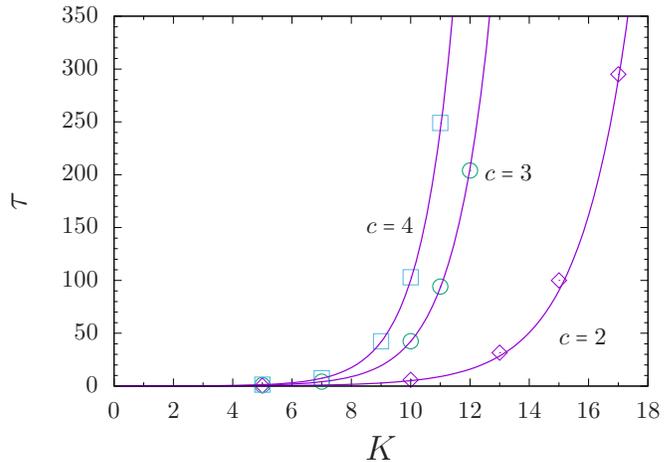}
	\caption{MTE for the processes given by $\text{A}  \xrightarrow{\lambda}  2\text{A}$ and $c\text{A}  \xrightarrow{\mu}  \varnothing$ vs $K$, for $\lambda=100$ and varying $\mu$. Symbols correspond to numerical simulations performed up to time $10^8$ and averaging over $1.5\times10^4$ realizations. Solid curves for $c=2,3,4$ are respectively obtained from Eqs.~(\ref{met1}), (\ref{met2}), and (\ref{met3}).}
	\label{fig:f2}
\end{figure}

%%%%%%%%%%%%%
One can also plot the MTE as a function of the carrying capacity $K$ keeping $\lambda$ fixed. As expected,
the MTE increases with the carrying capacity $K$ because the typical population size in the QSD increases, which naturally increases the entropic barrier to extinction. In Fig.~\ref{fig:f2} we
illustrate this and present our results for different values of $c$.
%%%%%%%%%%
%%%%%%%%%%%%%
%:2017-06-28 2:30

\subsection{Population abundance}\label{sec:abund}
In the previous subsection we have calculated the MTE for the case of $c=d$. If $d<c$, the extinction rate vanishes and the population survives. Biologically speaking,
extinction only occurs if the competition or death process is strong
enough, resulting in the death of all individuals that participate in the competition.
Otherwise, the population persists with a given abundance
that will depend on the parameters related to the individual interactions.

The deterministic dynamics of Eq.~(\ref{eq:tl}) is simple. Starting from a nonzero initial number of individuals, the population tends to the stable stationary state where the number of individuals equals $K$,  the carrying capacity of the medium. However, when intrinsic fluctuations are taken into account,  a stationary PDF of population sizes is reached as $t\to\infty$. This PDF is called the \textit{abundance}, and its mean approximately equals $K$ for $K\gg 1$. We remind the reader that for $d<c$, the absorbing state at $n=0$ can never be reached when starting from $n>0$ individuals.

We illustrate how to determine the stationary PDF, $P(n)$, for several prototypical examples with $d<c$.
If the reaction rates are polynomials in $n$,
it is sometimes more convenient to employ the probability generating function defined by~\cite{Ga90}
\begin{equation}
G(p,t)=\sum_{n=0}^{\infty}p^{n}P(n,t),\label{eq:dG}
\end{equation}
instead of using the WKB method, when dealing with the stationary solution of the master equation.
Here $p$ is an auxiliary variable, which is conjugate to the number
of particles~\cite{elgart2004rare}, and normalization of $P(n,t)$ implies that $G(p=1,t)=1$. Once $G(p,t)$ is known, the PDF is given by the Taylor
coefficients
\begin{equation}
P(n,t)=\frac{1}{n!}\left[\frac{\partial^{n}G(p,t)}{\partial p^{n}}\right]_{p=0}.\label{Pnt}
\end{equation}

Multiplying the master equation~(\ref{eq:me2}) by $p^{n}$,
summing over $n$, and renaming the index
of summation, we find after some algebra the evolution equation for the probability generating function,
\begin{equation}
\frac{\partial G(p,t)}{\partial t} = \frac{\lambda}{b!}p^{b}(p^{a}-1)\frac{\partial^{b}G}{\partial p^{b}} + \frac{\mu}{c!}p^{c-d}(1-p^{d})\frac{\partial^{c}G}{\partial p^{c}}.\label{eq:G}
\end{equation}
The solution of this evolution equation allows us to find the PDF for the general model described by Eqs.~(\ref{eq:1}).

We focus on the case of $a=b=1$, corresponding to the theta-logistic model described by Eqs.~(\ref{eq:rec1}). Then for the stationary solution,
$\partial_t G=0$, Eq.~(\ref{eq:G}) reads
\begin{equation}
\frac{d ^{c}G(p)}{d p^c}=\frac{\lambda c!}{\mu}p^{1-c+d}\frac{1-p}{1-p^d}\frac{d G(p)}{d p},
\label{eq:Gss}
\end{equation}
where we assume that $d<c$. This is an ordinary differential equation of order $c$, which can be exactly solved for specific values of $c$.

\subsubsection{$c=2,d=1$}
For $c=2$ and $d=1$,  the corresponding individual interactions are given by $\text{A} \xrightarrow{\lambda} 2\text{A}$, $2\text{A}  \xrightarrow{\mu}\text{A}$. This is the simplest case and
Eq.~(\ref{eq:Gss}) turns into
\begin{equation}
\frac{d ^{2}G(p)}{d p^2}=\frac{2\lambda }{\mu}\frac{d G(p)}{d p}.
\label{eq:Gss1}
\end{equation}
The boundary conditions for this equation are $G(p=1)=1$ and $G(p=0)=0$. The latter implies that the extinction probability vanishes, $P(n=0)=G(p=0)=0$, provided that we start from $n>0$ individuals. The solution of Eq.~(\ref{eq:Gss1}) with these boundary conditions reads
\begin{equation}
G(p)=\frac{e^{2\lambda p/\mu}-1}{e^{2\lambda /\mu}-1}.
\label{eq:Gss10}
\end{equation}
Expanding the term $e^{2\lambda p/\mu}$ around $p=0$ and comparing the result with Eq. (\ref{eq:dG}), we finally find
\begin{equation}
P(n)=\frac{K^n}{n!(e^{K}-1)},
\label{eq:Pss1}
\end{equation}
where, from Eq.~(\ref{rK}), $K=2\lambda /\mu$.  The mean value of the population in the steady state can be found using Eq.~(\ref{eq:Pss1}),
\begin{equation}
\langle n\rangle =\sum _{n=1}^{\infty}nP(n)=K\frac{e^K}{e^K -1},
\label{n1}
\end{equation}
which approaches $K$ when $K\gg 1$.

\subsubsection{$c=3$}
When $c=3$, the condition $0<d<c$ provides two possibilities, $d=2$ or $d=1$.

We consider first the case $c=3,d=2$ which corresponds to $\text{A} \xrightarrow{\lambda} 2\text{A}$, $3\text{A}  \xrightarrow{\mu}\text{A}$.
Then Eq.~(\ref{eq:Gss}) becomes
\begin{equation}
(1+p)\frac{d ^{3}G(p)}{d p^3}=\frac{6\lambda }{\mu}\frac{d G(p)}{d p},
\label{eq:Gss2}
\end{equation}
which can be solved using the boundary conditions as in the previous case $G(0)=0$, $G(1)=1$ and an additional \textit{self-generated} boundary condition that arises from the fact that Eq.~(\ref{eq:Gss2}) is singular at $p=-1$~\cite{Assaf2006stm,assaf2007spectral,assaf2010large}. Since $G(p)$ must be analytic everywhere, we require from (\ref{eq:Gss2}) that $G'(p=-1)=0$, where the prime denotes differentiation with respect to $p$.  The solution of Eq.~(\ref{eq:Gss2}) with these boundary conditions is
\begin{equation}
G(p)=\frac{(1+p)I_2 \left(2\sqrt{2(1+p)}K\right)-I_2 \left(2\sqrt{2}K\right)}{2I_2 \left(4K\right)- I_2 \left(2\sqrt{2}K\right)},
\label{eq:Gss2p}
\end{equation}
where $I_\alpha(\cdot)$ are the modified Bessel functions of order $\alpha$, and from Eq.~(\ref{rK}), $K=\sqrt{3\lambda /\mu}$. Expanding this solution in the vicinity of $p=0$ and comparing with Eq.~(\ref{eq:dG}), we arrive after some algebra at the solution
\begin{equation}
P(n)=\frac{\frac{1}{n!}\left(\sqrt{2}K\right)^n I_{2-n}\left(2\sqrt{2}K\right)}{2I_2 \left(4K\right)- I_2 \left(2\sqrt{2}K\right)}.
\label{eq:Pn2}
\end{equation}
The mean value follows immediately,
  \begin{equation}
  \langle n\rangle =\sum _{n=1}^{\infty}nP(n)=K\frac{2I_1(4K)}{2I_2 \left(4K\right)- I_2 \left(2\sqrt{2}K\right)},
  \label{n2}
  \end{equation}
which again converges to $K$ as $K\to\infty$.

\begin{figure}[htbp]
	\includegraphics[width=\hsize]{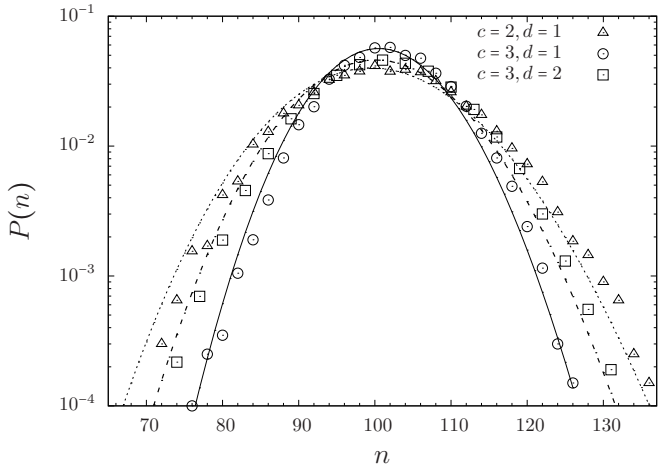}
	\caption{Population abundance $P(n)$ as a function of $n$
	for $K=100$.
	Theoretical results (lines) from Eqs.~(\ref{eq:Pss1}), (\ref{eq:Pn2}), and (\ref{eq:Pn3}) are compared with numerical simulations (symbols), where we have performed $10^4$ realizations from an initial population equal to $K$ up to $10^7$ time steps.}
	\label{fig:f3}
\end{figure}

\begin{figure}[htbp]
	\includegraphics[width=\hsize]{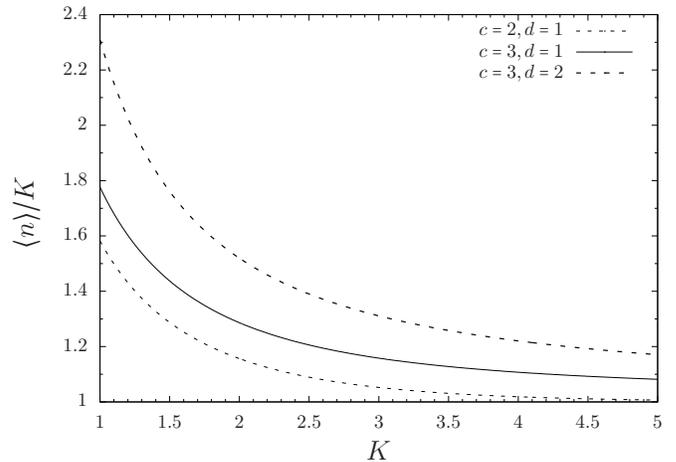}
	\caption{Mean population abundance $\langle n\rangle$ divided by $K$. Shown are theoretical results from Eqs.~(\ref{n1}), (\ref{n2}), and (\ref{n3}). The mean population abundance approaches $K$ as $K\to\infty$.}
	\label{fig:f4}
\end{figure}

Finally we consider the second possibility where $c=3$ and $d=1$, which corresponds to $\text{A} \xrightarrow{\lambda} 2\text{A}$, $3\text{A}  \xrightarrow{\mu} 2\text{A}$.
Equation (\ref{eq:Gss}) reduces to
\begin{equation}
p\frac{d ^{3}G(p)}{d p^3}=K^2\frac{d G(p)}{d p},
\label{eq:Gss3}
\end{equation}
where, from Eq.~(\ref{rK}), $K=\sqrt{6\lambda /\mu}$.
This equation can be solved with the  boundary conditions $G(0)=0$, $G(1)=1$, and the self-generated boundary condition, $G'(p=0)=0$, which cures the singularity of Eq.~(\ref{eq:Gss3}) at $p=0$. The solution of (\ref{eq:Gss3}) with these boundary conditions reads
\begin{equation}
G(p)=\frac{pI_2 \left(2K\sqrt{p}\right)}{I_2 \left(2K\right)}.
\label{eq:Gss3p}
\end{equation}
Expanding the numerator in the vicinity of $p=0$ and comparing with (\ref{eq:dG}) we obtain for $n>1$
\begin{equation}
P(n)=\frac{K^{2n-2}}{n!(n-2)!I_2 \left(2K\right)}.
\label{eq:Pn3}
\end{equation}
The mean population in this case is
 \begin{equation}
 \langle n\rangle =\sum _{n=1}^{\infty}nP(n)=K\frac{I_1(2K)}{I_2 \left(2K\right)}.
 \label{n3}
 \end{equation}
In Fig.~\ref{fig:f3} we plot the population abundance obtained in the three cases described above, Eqs.~(\ref{eq:Pss1}), (\ref{eq:Pn2}), and (\ref{eq:Pn3}), and compare our analytical predictions with numerical simulations.
%%%%%%%%%%
%%%%%%%%%%%%%
Finally, in Fig.~\ref{fig:f4} we plot the mean population value for different values of $K$ obtained from Eqs.~(\ref{n1}), (\ref{n2}), and (\ref{n3}).
%%%%%%%%%%%%%
%
As expected, the mean population value tends to the deterministic value as the carrying capacity $K$ increases. The mean value of the PDF deviates noticeably from the deterministic steady state for $K=O(1)$, when relative fluctuations are large.
In summary, we find that the microscopic details of individual interactions strongly affect the population abundance in the steady state and its mean value.

\section{Stochastic dynamics of the theta-logistic equation. Three-reactions model}\label{sec:three}

In this section we add a linear death reaction, $\text{A} \xrightarrow{\gamma}\emptyset$,
which describes a natural death process with rate $\gamma$,
to the previous model, Eqs.~(\ref{eq:rec1}).
The presence of this linear death process
ensures that the population
always becomes extinct, regardless of the value of $d$.
Considering the new transition probability, $W(n,-1)=\gamma n$,
in addition to the transitions (\ref{eq:trc1}), the
master equation takes the form
\begin{multline}
\frac{\partial P(n,t)}{\partial t}  =  \lambda(n-1)P(n-1,t)+\gamma(n+1)P(n+1,t)\\
  +  \frac{\mu}{c!}\frac{(n+d)!}{(n+d-c)!}P(n+d,t) \\
  -  \left[\lambda n+\frac{\mu}{c!}\frac{n!}{(n-c)!}+\gamma n\right]P(n,t).\label{eq:mec1c3}
\end{multline}
Rescaling the time $t\to \gamma t$, the dimensionless
transition probabilities now become $w_{1}=R_{0}q$, $u_{1}=0,$ $w_{-1}=q,$
$u_{-1}=0,$ $w_{-d}=R_{0}q^{c}/c!$, and $u_{-d}=-R_{0}q^{c-1}/(2(c-2)!)$,
where
\begin{equation}
R_{0}=\frac{\lambda}{\gamma},\quad N=(\lambda/\mu)^{\frac{1}{c-1}}.
\end{equation}
The Hamiltonian in this case can be calculated from Eq.~(\ref{eq:H}),
\begin{equation}
H(q,p)=R_{0}q\left(e^{p}-1\right)+R_{0}\frac{q^{c}}{c!}\left(e^{-dp}-1\right)+q\left(e^{-p}-1\right).\label{eq:Hc3}
\end{equation}
The mean-field equation can be found from the Hamilton's equation
$\dot{q}=\partial_{p}H$ evaluated at $p=0$. By setting $n=qN$, we recover the theta-logistic equation (\ref{eq:tl}). However, here the intrinsic growth rate and the carrying capacity, which are defined in terms of the parameters characterizing the individual interactions, are different from Eq.~(\ref{rK}) and are given by
\begin{equation}
r=\lambda-\gamma,\quad K=\left[\frac{c!(\lambda-\gamma)}{d\mu}\right]^{\frac{1}{c-1}}.
\end{equation}
Since the intrinsic growth rate must be positive, we must have $\lambda>\gamma$, such that the basic reproductive rate $R_{0}$ is greater than $1$.
Here, the population density in the steady state is given by
$q_{*}=K/N=[c!(R_{0}-1)/(dR_{0})]^{\frac{1}{c-1}}$.

To find the MTE for this model, we need to determine the nontrivial zero-energy trajectory, the optimal path to extinction, of the Hamiltonian (\ref{eq:Hc3}),  which reads
\begin{equation}
q_{a}(p)=\left\{ \frac{c![R_{0}q\left(e^{p}-1\right)-1+e^{-p}]}{R_{0}\left(1-e^{-dp}\right)}\right\} ^{\frac{1}{c-1}},\label{eq:qac3}
\end{equation}
where $p_{f}=p(q=0)=-\ln R_{0}$.

As before, we have to determine the WKB solution for the QSD and match it with a recursive solution in their joint region of validity $N^{-1}\ll q\ll 1$ to find the MTE.
In the limit of $q\ll 1$, the action function $S(q)$ can be calculated by expanding $q_a(p)$ in the vicinity of $p_f$. Defining  $\epsilon=p-p_{f}\ll1$, substituting it into Eq.~(\ref{eq:qac3}), and expanding in a
power series to the lowest
order of $\epsilon$, we find $p_{a}(q)\simeq-\ln R_{0}+O(q)$. As a result, using Eq.~(\ref{eq:sq}),
we have $S(q)\simeq-\ln (R_{0}q)+O(q^{2})$. Consequently, the $q\ll 1$ asymptote of the WKB solution for the QSD, given by Eq.~(\ref{eq:pwkb2}), becomes
\begin{equation}
P_{\text{WKB}}(n)  \simeq  \sqrt{\frac{S''(q_{*})}{2\pi N}}\frac{Nq_{*}}{n}e^{-N\Delta S-\Delta\phi}R_{0}^{n},\label{eq:qsdc3}
\end{equation}
where the quantities $\Delta S$ and $\Delta\phi$ are given by
\begin{equation}
\Delta S  =  \int_{-\infty}^{0}q_{a}(p)dp = \int_{1/R_{0}}^{1}\left[c!\frac{(1-z)(z-R_{0}^{-1})}{(1-z^{d})z^{c-d}}\right]^{\frac{1}{c-1}}dz,
\label{eq:Dsccas3}
\end{equation}
and
\begin{equation}
\Delta\phi  =  \frac{1}{2}\ln\left(\frac{2dR_{0}^{c}(R_{0}-1)}{(R_{0}^{d}-1)[R_{0}(d+1)-d+1]}\right).\\
\end{equation}

Now all that is left is to calculate the recursive solution for the QSD for small values of $n$ and match it
to the above solution. Here, however, we take a different approach, because of the presence of the linear death rate, and linearize the transition rates in the vicinity of $n=0$, in the spirit of~\cite{Assaf2010ems}: $W(n,1)\simeq\lambda n,$ $W(n,-d)\simeq0$, and $W(n,-1)\simeq\gamma n$. As a result, for $n\ll N$, the quasi-stationary master equation becomes
\begin{equation}
P(n+1)\simeq(R_{0}+1)\frac{n}{n+1}P(n)-R_{0}\frac{n-1}{n+1}P(n-1).\label{eq:rP}
\end{equation}
Defining $f(n)=nP(n)$, the stationary solution of Eq.~(\ref{eq:rP}) reads
$f(n)\simeq C_{0}+C_{1}R_{0}^{n}$ \cite{Assaf2010ems}. To obtain the unknown constants
$C_{0}$ and $C_{1}$, we need to specify two boundary conditions.
One boundary condition is that $f(0)=0$, which allows us to express both $C_{0}$ and $C_{1}$
in terms of $f(1)$. As a result we find
\begin{equation}
P(n)\simeq\frac{R_{0}^{n}-1}{n(R_{0}-1)}f(1).\label{eq:Pnrec}
\end{equation}
We now relate $f(1)$ to the MTE. Equation (\ref{eq:met1}) implies that
$\tau^{-1}=W(d,-d)P(d)+W(1,-1)P(1)$. For small $n$,
one has $W(d,-d)\simeq 0$, and thus $\tau^{-1}\simeq\gamma f(1)$.
As a result, Eq.~(\ref{eq:Pnrec}) becomes
\begin{equation}
P(n)\simeq\frac{R_{0}^{n}-1}{n(R_{0}-1)\gamma\tau}.\label{eq:Pnrec2}
\end{equation}

\begin{figure}[htbp]
\includegraphics[width=\hsize]{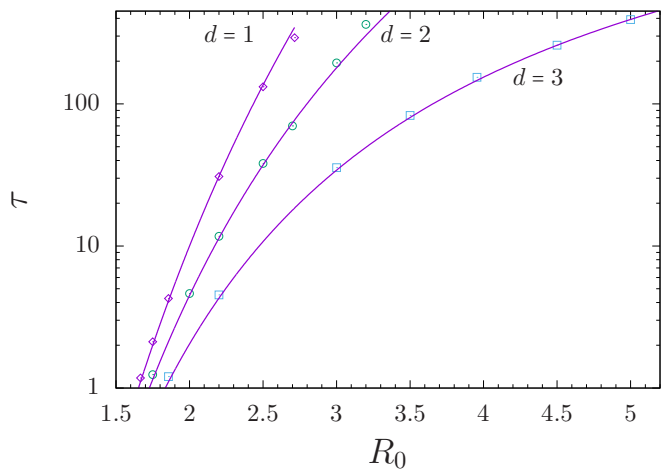}
\caption{Plot of the MTE for three different reaction schemes with different values of $d$. The parameters are: $r=600$, $c=3$, and $K=20$. Symbols correspond to numerical simulations performed up to time $10^9$ and averaging over $10^4$ realizations. Solid curves are obtained from Eq.~(\ref{eq:MTEcas3}) for $d=1,2,3$.}
\label{fig:f5}
\end{figure}

Matching the $n\gg 1$ asymptote of this recursive solution to the WKB asymptote, Eq.~(\ref{eq:qsdc3}), in their joint region of validity, $1\ll n\ll N$, we obtain the MTE
\begin{equation}
\tau=\sqrt{\frac{2\pi}{N(c-1)}}\frac{R_{0}^{c/2} e^{N\Delta S}}{\gamma\sqrt{(R_{0}-1)(R_{0}^{d}-1)}}\left[\frac{R_{0}d^{c}}{c!\left(R_{0}-1\right)^{c}}\right]^{\frac{1}{2(c-1)}}
\label{eq:MTEcas3}
\end{equation}
where $\Delta S$ is given by (\ref{eq:Dsccas3}) and we have made
use of $S''(q_{*})=[dq_{a}(p)/dp]_{p=0}^{-1}$.
The behavior of $\tau$ in terms of the basic reproductive rate $R_0$ for different values of $d$ is shown in Fig.~\ref{fig:f5}.
%%%%%%%%%%
%%%%%%%%%%%%%
The MTE increases with $R_0$, and it decreases with $d$ for fixed $R_0$, as expected. The figure illustrates that the MTE is highly sensitive to the details of the  microscopic reaction scheme.

\section{Stochastic dynamics of von Bertalanffy's equation}\label{sec:doom}
\subsection{Mean-field dynamics}
As a final example, we consider a special case of Eq.~(\ref{eq:1}) with $c=d=a=1$ and $b>1$. The reaction
scheme reduces to
\begin{subequations}\label{Bertalan}
\begin{align}
b\text{A} & \xrightarrow{\lambda} (b+1)\text{A}\\
\text{A} & \xrightarrow{\mu} \varnothing.
\label{vB}
\end{align}
\end{subequations}
Equation (\ref{eq:mfe1}) implies that in this case
the mean field equation
reduces to
\begin{equation}
\frac{dn}{dt}=\frac{\lambda}{b!}n^{b}-\mu n,\label{eq:mfe2}
\end{equation}
which is known as the von Bertalanffy's equation in population dynamics \cite{Be57,Be60}. It has
two steady states: $n=0$ and $n_*=(\mu b!/\lambda)^{1/(b-1)}$.
Since $b>1$, the first term of the right hand side of \eqref{eq:mfe2}
is nonlinear. Linearizing Eq.~\eqref{eq:mfe2} around the
steady states, we find that $n=0$ is an attractor, whereas $n=n_*$ is a repeller.  Equation (\ref{eq:mfe2}) can be integrated exactly,
\begin{equation}
n(t)=\frac{1}{\left[\frac{\lambda}{b!\mu}-\left(\frac{\lambda}{b!\mu}-n_0^{1-b}\right)e^{(b-1)\mu t}\right]^{\frac{1}{b-1}}},
\label{soldd2}
\end{equation}
where $n_0=n(t=0)$. Depending on the initial condition and in the absence of fluctuations, this solution may represent either an extinction scenario or an unlimited population growth. If $n_0<n_*$, then the denominator of (\ref{soldd2}) is always positive and the population density tends to $0$ as $t\to \infty$. If on the other hand $n_0>n_*$,
then the denominator becomes zero at a finite time, $t_c$, which is known in the ecological
literature as \textit{doomsday} \cite{Fo60,Pa15}. Here, $t_c$ satisfies
\begin{equation}
t_{c}=-\frac{1}{\mu(b-1)}\ln\left(1-n_{0}^{1-b}\frac{b!\mu}{\lambda}\right).
\end{equation}
However, if we take internal fluctuations into account,
extinction can occur with a \textit{non-zero} probability, even if the population starts at $n_0>n_*=(\mu b!/\lambda)^{1/(b-1)}$.
We determine the extinction probability in the next section.

\subsection{Extinction probability}
We write the birth and death rates, respectively, of system~(\ref{Bertalan}) as
\begin{equation}\label{rece}
W^{+}_n=\frac{\lambda}{b!}\frac{n!}{(n-b)!}, \quad W^{-}_n=\mu n.
\end{equation}
We are interested in calculating the probability of extinction of a population, starting from $n>n_*$ and obeying the interactions given by (\ref{Bertalan}). As previously, we denote the typical system size by $N$, such that $N\equiv n_{*}=(\mu b!/\lambda)^{1/(b-1)}$, and $Q(n)$ is
the extinction probability starting from $n$ individuals.
When $n<N$, this probability is close to unity, since there is a mean-field flow towards $0$. However, starting from $n>N$, the probability of extinction decreases as $N$ increases, since $n=N$ is an unstable fixed point.

The recursion equation for $Q(n)$ is given by\cite{antal2006fixation,mobilia2010fixation,assafmobilia2010}
\begin{equation}\label{qn}
W^{+}_n Q(n+1)+W^{-}_n Q(n-1)-[W^{+}_n +W^{-}_n]Q(n)=0.
\end{equation}
It reflects the fact the probability of extinction starting from $n$ individuals
is the probability of extinction starting from $n+1$ individuals,
multiplied by
the probability to reach state $n+1$ from state $n$, plus the probability of
extinction starting from $n-1$ individuals, multiplied by the probability
to reach state $n-1$ from state $n$.
Equation (\ref{qn}) is supplemented with the boundary conditions,
see, e.g., Refs.\ \cite{antal2006fixation,mobilia2010fixation,assafmobilia2010},
\begin{equation}
Q(0)=1,\quad Q(\infty)=0.
\label{bcQ}
\end{equation}
Let
\begin{equation}
R(n)=Q(n+1)-Q(n).
\label{RQ}
\end{equation}
The equation for $R(n)$ reads
\begin{equation}
W^{+}_n R(n)-W^{-}_n R(n-1)=0,
\end{equation}
which can be expressed as
\begin{equation}\label{err}
(n-1)(n-2)\dots (n-b+1)R(n)=R(n-1)N^{b-1}
\end{equation}
by virtue of Eq.~(\ref{rece}).
This equation self-generates the boundary conditions for $R(n)$,
which are found to be $R(0)=R(1)=R(2)=\dots =R(b-2)=0$.
By iterating Eq.~(\ref{err}) and taking into account the above boundary conditions,
we arrive at the solution
\begin{equation}\label{err1}
R(n)=R(b-1)N^{(b-1)(n-b)}\prod_{j=b}^{n}\frac{(j-b)!}{(j-1)!}.
\end{equation}
Using Eq.~(\ref{RQ}) and the boundary conditions for $Q(n)$, we obtain
\begin{equation}
Q(n)=1+\sum_{m=b-1}^{n-1}R(m).
\label{QR}
\end{equation}
Since $Q(\infty)=0$, we have
\begin{equation}
\sum_{m=b-1}^{\infty}R(m)=-1.
\label{cR}
\end{equation}
This condition allows us to find the unknown constant $R(b-1)$
in (\ref{err1}). Using Eqs.~(\ref{err1}), (\ref{QR}), and (\ref{cR}), we finally have
\begin{equation}
Q(n)=1-\frac{\sum_{m=b-1}^{n-1}N^{(b-1)(m-b+1)}\prod_{j=b}^{m}\frac{(j-b)!}{(j-1)!}}{\sum_{n=b-1}^{\infty}N^{(b-1)(n-b)}\prod_{j=b}^{n}\frac{(j-b)!}{(j-1)!}}.
\label{Qnf}
\end{equation}
This is an exact expression valid for any value of $b$.

Let us evaluate this expression for several specific values of $b$.
For example, for $b=2$, Eq.~(\ref{Qnf}) becomes
\begin{equation}
Q(n)=1-\frac{\Gamma (n-1,N)}{\Gamma (n-1)},
\label{Qn2}
\end{equation}
where $\Gamma (,)$ is the incomplete Gamma function.
For $b=3$ we have $Q(1)=Q(2)=1$, and for $n\geq 3$ the solution reads
\begin{equation}
Q(n)=\frac{N^{2n-4}(n-1)(n-2)}
{\Gamma(n)^{2}I_{2}(2N)}{}_{1}F_{2}\left(1;n,n-2;N^2\right),
\label{Qn3}
\end{equation}
where ${}_{p}F_{q}(\cdot)$ is the generalized hypergeometric function.
Finally, for $b=4$ we have $Q(1)=Q(2)=Q(3)=1$, and
for $n\geq 4$ the solution reads
%%%%%%%%%%
\begin{figure}[htbp]
	\includegraphics[width=\hsize]{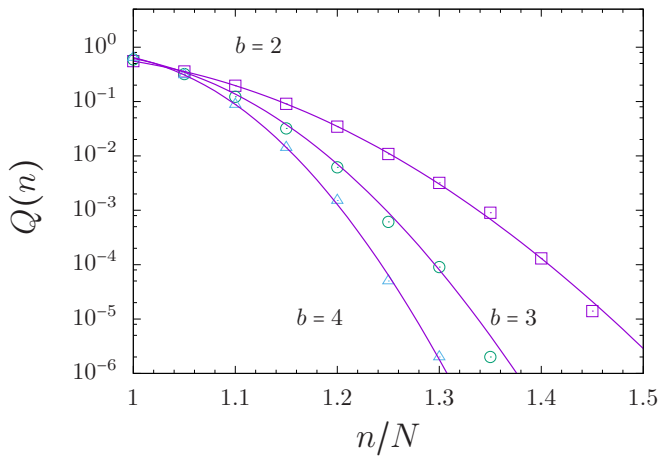}
	\caption{Probability of extinction Q(n) starting from $n$ individuals. We have considered here $\mu =6$, $N=100$ and $b=2,3,4$. The symbols correspond to numerical simulations, and the solid curves are obtained from the exact solutions Eqs.~(\ref{Qn2}), (\ref{Qn3}), and (\ref{Qn4}). The number of events considered in the simulations is $10^7$.}
	\label{fig:f6}
\end{figure}
%%%%%%%%%%%%%
\begin{equation}
Q(n)=\frac{N^{3n}{}_{1}F_{3}\left(1;n,n-1,n-2;N^3\right)}{(n-1)!(n-2)!(n-3)!\theta(N)},
 \label{Qn4}
 \end{equation}
where
\begin{equation}
\theta(N)=\sum_{n=3}^{\infty}\frac{N^{3n}}{(n-1)!(n-2)!(n-3)!}.
\end{equation}
In Fig.~\ref{fig:f6} we plot the exact solutions (\ref{Qn2}), (\ref{Qn3}), and (\ref{Qn4}), which are in excellent agreement with numerical simulations.

\section{Conclusions}\label{sec:concl}
We have constructed several individual-based models that give rise to the generalized logistic model at the mean-field deterministic level. Importantly, and unlike previous studies that used ad hoc effective rates, our microscopic models, based on multi-step reaction schemes for birth and death processes, allow us
to interpret the parameters of the deterministic model in terms of the interactions between individuals. For the different deterministic models that have been widely used in the literature, we have studied their corresponding microscopic analogs and have
taken into account the effect of internal demographic fluctuations on the long-time dynamics and the conditions for extinction. In particular, when extinction takes place, we have analytically determined the mean time to extinction using the WKB method. To the best of our knowledge, this is the first derivation of an analytical expression of the MTE for the generalized logistic model.
For those models that do not exhibit extinction, we have analytically derived the stationary population abundance distribution, using the probability generating function formalism. Importantly we have found that microscopic models that display different long-time behaviors, namely extinction or persistence, obey nevertheless the same mean-field equation.
We have also provided novel results for runaway systems, i.e., populations that undergo a doomsday scenario. We have shown that such systems can undergo extinction events, even under doomsday conditions, due to internal fluctuations. We have derived analytical expressions for the extinction probability of such populations. All our theoretical predictions have been verified by numerical simulations. We expect that our identification of the macroscopic parameters widely used in ecology in terms of the actual rates of individual interactions and our simple and analytically amenable expressions for the mean time to extinction, population abundance and extinction probability will have impact in the fields of theoretical ecology and biodiversity.

\begin{acknowledgments}
This research has been supported (VM, DC) the Ministerio de Econom\'{\i}a y Competitividad under Grant No.CGL2016-78156-C2-2-R. VM acknowledges the hospitality of the Department of Chemistry and Biochemistry of the University of California, San Diego where part of this work was developed.
\end{acknowledgments}

%\vfill
%\bibliographystyle{abbrv}
\bibliography{gen2}

\section*{Appendix}
\renewcommand{\theequation}{A\arabic{equation}}
\setcounter{equation}{0}
In this appendix we briefly present the WKB calculation for the (quasi)stationary distribution (QSD)\cite{Dykman1994lfo,kessler2007extinction,meerson2008noise,escudero2009switching,Assaf2010ems,AssafMeerson2017}. Our starting point is the (quasi)stationary master equation~(\ref{eq:mec1}) with $\partial_tP(n,t)=0$. Employing the WKB ansatz for the QSD, $P(n)\equiv P(Nq)$~\cite{Dykman1994lfo},
\begin{equation}
P_{\text{WKB}}(q)\equiv Ae^{-NS(q)-S_{1}(q)},
\label{Pwkb}
\end{equation}
and substituting it into the quasi-stationary master equation, we can determine
the functions $S(q)$ and $S_1(q)$ order by order for $N\gg 1$.
In the leading order we find the action function to be
\begin{equation}
S(q)=\int p_{a}(q)dq.\label{eq:sq}
\end{equation}
Here $q_{a}(p)$ or $p_{a}(q)$ define the nontrivial optimal path to extinction,
$H=0$, where the Hamiltonian satisfies
\begin{equation}
H(q,p)=\sum_{r}w_{r}(q)\left(e^{rp}-1\right),\label{eq:H}
\end{equation}
and $p=S'(q)$ is the associated momentum \cite{Dykman1994lfo}.
Note that the mean-field dynamics, Eq.~(\ref{eq:tl}), can be found
by writing the Hamilton's equation
$\dot{q}=\partial_{p}H$ along the deterministic (noise free) path
$p=0$
\cite{Dykman1994lfo,kessler2007extinction,meerson2008noise,escudero2009switching,Assaf2010ems,AssafMeerson2017}.

In the subleading $N^{-1}$ order, we find
\begin{equation}
S_{1}'(q)=\left.\frac{\frac{1}{2}S''(q)\partial_{pp}H+\partial_{qp}H-\sum_{r}u_{r}(q)\left(e^{rp}-1\right)}{\partial_{p}H}\right\rvert_{p=p_{a}(q)}.\label{eq:s1p}
\end{equation}

The constant $A$ in Eq.~(\ref{Pwkb})
can be obtained by normalizing the QSD to unity in the Gaussian vicinity of
the deterministic stable state $q_{*}$. Expanding the QSD near
$q=q_{*}$ up to second order, the Gaussian limit, integrating over $q$,
and equating to one, we find that the constant $A$ has the form
\begin{equation}
A=\sqrt{\frac{S''(q_{*})}{2\pi N}}e^{NS(q_{*})+S_{1}(q_{*})}.
\end{equation}
Substituting this expression into Eq.~(\ref{Pwkb}), we obtain
\begin{equation}
P_{\text{WKB}}(q)=\sqrt{\frac{S''(q_{*})}{2\pi N}}e^{NS(q_{*})+S_{1}(q_{*})-NS(q)-S_{1}(q)}.\label{eq:qs1}
\end{equation}
Defining $S_{1}(q)=\phi(q)+\ln q$, the QSD takes the final
form
\begin{equation}
P_{\text{WKB}}(q)=\sqrt{\frac{S''(q_{*})}{2\pi N}}\frac{q_{*}}{q}e^{N[S(q_{*})-S(q)]}e^{\phi(q_{*})-\phi(q)},\label{eq:qs2}
\end{equation}
where $S(q)$ is defined in (\ref{eq:sq}),
\begin{equation}
\phi(q)=S_{1}(q)-\ln q=\int\varLambda[q,p=p_{a}(q)]dq,\label{eq:f1}
\end{equation}
and
\begin{multline}
\varLambda(q,p)=\frac{\partial_{qp}H(q,p)+\frac{1}{2}\partial_{pp}H(q,p)p_{a}'(q)}{\partial_{p}H(q,p)}\\
-\frac{\sum_{r}u_{r}(q)\left(e^{rp}-1\right)}{\partial_{p}H(q,p)}-\frac{1}{q}.
\label{Lambda}
\end{multline}
This QSD is valid as long as $q\gg N^{-1}$, or $n\gg 1$. In order to find the QSD for  $n=O(1)$ one has to solve a recursive equation, see main text.

\end{document}